\documentclass[10pt,showpacs,preprintnumbers,amsmath,amssymb]{revtex4}
\usepackage{units}
\usepackage{bbold}
\usepackage{graphicx}
\usepackage{dcolumn}
\usepackage{bm}
\usepackage{slashed}

\begin{document}

\title{Axisymmmetric empty space: light propagation, orbits and dark matter}
\vspace{2cm}
\author{Sergio Giardino}
\email{giardino@ime.unicamp.br}
\affiliation{ Instituto de Matem\'{a}tica, Estat\'{i}stica e
Computa\c{c}\~{a}o Cient\'{i}fica, Universidade Estadual de Campinas\\
Rua S\'{e}rgio Buarque de Holanda 651, 13083-859, Campinas, SP,
Brazil}

\begin{abstract}
\noindent This study presents a axisymmetric solution of the Einstein equations for empty space. The geometry is studied by determining its Petrov classification and  Killing vectors. Light propagation, orbital motion and asymptotic and Newtonian limits are also studied. Additionally, cosmological applications of the geometry as an alternative model for the inflationary universe and as a substitute for dark matter and quintessence are also outlined.\linebreak
\linebreak
\end{abstract}
\maketitle

\section{Introduction}
Axial symmetry, which comprises both spherical symmetry and cylindrical symmetry, is widely used to simplify a variety of physical problems.  Spherical  symmetry is the most preferred option  to simplify a physical problem, because its highly symmetrical content. In many cases, a cylindrically symmetric problem has a dynamic character associated with a rotation, and a limit without rotation is a spherically symmetric situation. One could suppose that such situation occurs in general relativity, where the  Schwarzschild solution is found in the rest limit of the rotating Kerr black hole. However, both have two Killing vectors, one time-like Killing vector and a axisymmetryc Killing vector, and then have the same symmetry. A similar situation occurs through a deformation of spherical symmetry. Another example occurs between the $AdS_5\times S^5$ and Lunin-Maldacena \cite{Lunin:2005jy} space-time solutions. Both spaces are ten-dimensional and consist of a five-dimensional anti-de Sitter space. The difference between them occurs in the five-dimensional spherical sector. In the Lunin-Maldacena solution this sector consists of a deformed sphere, whose deformation is parametrized. When the parameter is set to zero, the deformed sphere becomes the usual five dimensional sphere and consequently the Lunin-Maldacena space recovers $AdS_5\times S^5$ in this limit. Some other axisymmetric  space-times have been used in semi-classical string theory \cite{Giardino:2013sia,Giardino:2013rca}.

In general relativity, the Schwarzschild empty space is the simplest spherically symmetric solution, and probably the most important, and there are also many cilindrically symmetric solutions \cite{Stephani:2009ees}. The Weyl class, for example, has the general form 
\begin{equation}\label{weyl}
ds^2=-e^{2U}\,dt^2+e^{-2U}\big[e^{2V}\big(dr^2+dz^2\big)+r^2d\phi^2\big],
\end{equation}
with $U=U(r,\,z)$ and $V=V(r,\,z)$. These solutions have been known since the 1930s and have been applied to a variety of problems, like asymptotically flat space-times in terms of multipole expansions \cite{Sharif:2005ia,Backdahl:2005uz,HernandezPastora:2011gw} and $f(R)$ gravity \cite{Momeni:2009tk}. First attempts to build cylindrically symmetric solutions and develop generation techniques are reviewed in \cite{reina:1971tre}. Some of these older solutions, like the Lewis and the van Stockum solutions are not asymptotically flat, a property that we will consider in the solution reported here.  More recently, cylindrically symmetric solutions in general relativity have been studied, like black hole solutions \cite{Lemos:1994xp,Lemos:1995cm}, Brans-Dicke theory \cite{Baykal:2009vs} and black strings in Chern-Simons modified gravity \cite{Ahmedov:2010fz}.

In this article, we present a axisymmetric solution which is analogous to the spherically symmetric empty space. This means that the Einstein tensor, and consequently the Ricci tensor, vanish. The solution is not new, but the form in which it is presented here certainly is.  The solution is quite simple, but it also has some interesting features. The first feature is that it is not asymptotically flat in the region very far from the origin of the coordinate system, but it is flat in the region closely around the origin of the coordinate system. This local flatness is, of course, a property of every point of a differential manifold. However, the farther the distance from the center of the coordinate system, the higher the gravitational field. Accordingly, the Kretschsmann scalar presents a divergence in the $r\to \infty$ limit. This counter-intuitive fact suggests that there is some gravitational source at an infinite distance from the origin of the coordinate system. As the symmetry of the problem is axial, this source may be a ring, a cylinder or some revolution surface. The net effect is that every mass would be attracted to the far region of this universe. The similarity with inflationary cosmological models is possible, and then such a geometry may be an alternative to other modes that predict an expansion of the universe, like inflation, quintessence and cosmological constant \cite{Sami:2014faa,Sami:2013ssa}. From this standpoint, the role played by the cosmological constant or a scalar field can be changed by a point at infinity, which has a mass associated with it, such as the case of the Schwarzshild solution. On the other hand, there is a big difference between the solution presented and the Schwarzschild solution: there is no event horizon. This $r\to \infty$ point, if it is singular, may not violate the cosmic censorship hypothesis. As the point is located at infinity, it is not visible with or without an event horizon. Even if there was an event horizon, this would be located at infinity, and the result would be the same. Of course, these possibilities require careful analysis in order to be tested, but the solution presented herein seems simple enough to be a model for these theories. 

This article is organized as follows: in Section \ref{S2} we present the solution, its Petrov classification and its Killing vectors and we calculate the Kretschmann scalar. In Section \ref{S3}, we study light propagation in  space time. In Section \ref{S4} we study the existence of orbital planetary motion in the metric. In Section \ref{S5} we propose a Newtonian limit to the geometry and  Section \ref{S6} the author's conclusions are presented.

\section{The metric \label{S2}}
We seek an empty space solution, so that the Einstein tensor 
\begin{equation}
G_{\mu\nu}=R_{\mu\nu}-\frac{1}{2}\,g_{\mu\nu}R
\end{equation}
 is identical to zero, which is equivalent to $R_{\mu\nu}=0$. The axisymmetric ansatz is
\begin{equation}\label{metric}
ds^2=-\frac{u}{v^2}\,c^2dt^2+\frac{1}{v^4}dr^2+\frac{1}{uv^2}dz^2+r^2d\phi^2,
\end{equation}
where $u=u(z)$ and $v=v(r)$ and $c$ is the speed of light. The $R_{rr}$ component of the Ricci tensor provides the equation for $v$, namely
\begin{equation}\label{Rrr}
 r\,v_{rr}-v_r=0.
\end{equation}
Solving (\ref{Rrr}) and using its solution in the metric, we discover that the solutions of Einstein equations for the ansatz involve
\begin{equation}
u=4\,C_1\,C_2\,z^2+C_3\,z+C_4\qquad\mbox{and}\qquad v=C_1+C_2\,r^2,
\end{equation}
where $C_{i=1,2,3,4}$ are integration constants. We immediately see that the metric is almost flat around the point $z=r=0$  when $C_1=C_4=1$. Local flateness is a property of every point of the manifold, and hence every property of this point can be extended to the whole space by a coordinate translation.

In order to characterize the metric (\ref{metric}), we need to study its characteristics, which do not depend on the coordinate system. The symmetries can be determined through the Killing vectors, whose components satisfy
\begin{equation}
g_{\lambda(\mu}\xi^\kappa_{,\nu)}+g_{\mu\nu,\kappa}\xi^\kappa=0,
\end{equation}
where the comma means a derivative and $\xi^\kappa$ are the Killing vector components. In this solution, $\xi^r=\xi^z=0$ and $\xi^t$ and $\xi^\phi$ are constants. This results confirms the expectation that the solution is axisymmetric, as the Killing vector in the angular direction commutes with the null vector of the other spatial. 

Another important characterization of the solution is obtained through the Petrov classification. Defining the tetrad
\begin{equation}
u_\mu=\big(\sqrt{g_{00}},\,0,\,0,\,0\big),\qquad r_\mu=\big(0,\,\sqrt{g_{11}},\,0,\,0\big)\qquad z_\mu=\big(0,\,0,\,\sqrt{g_{22}},\,0\big)\qquad\mbox{and}\qquad \phi_\mu=\big(0,\,0,\,0\,\sqrt{g_{33}}\big),
\end{equation}
we build the null tetrad $e_\mu=\big(l_\mu,\,n_\mu,\,m_\mu,\,\overline{m}_\mu\big)$
\begin{equation}
l_\mu=\frac{1}{\sqrt{2}}\big(u_\mu+z_\mu\big)\qquad n_\mu=\frac{1}{\sqrt{2}}\big(u_\mu-z_\mu\big)\qquad m_\mu=\frac{1}{\sqrt{2}}\big(r_\mu+i\,\phi_\mu\big)\qquad\mbox{and}\qquad \overline{m}_\mu=\frac{1}{\sqrt{2}}\big(r_\mu-i\,\phi_\mu\big),
\end{equation}
which satisfies
\begin{equation}
l_\mu n^\mu= -\, m_\mu\overline{m}^\mu=1,\qquad e_\mu e^\mu= l_\mu m^\mu=l_\mu \overline{m}^\mu=m_\mu n^\mu=\overline{m}_\mu n^\mu=0\qquad\mbox{and}\qquad
 g_{\mu\nu}= l_\mu n_\mu + n_\mu l_\nu - \overline{m}_\mu m_\nu - m_\mu \overline{m}_\nu.\nonumber
\end{equation}
Thus, using the Weyl tensor $C_{\mu\nu\kappa\lambda}$ we calculate the Weyl scalars
\begin{eqnarray}
&&\Psi_0\,=\,C_{\mu\nu\kappa\lambda}\,l^\mu\, m^\nu \,l^\kappa\, m^\lambda\,=\,0,\\
&&\Psi_1\,=\,C_{\mu\nu\kappa\lambda}\,l^\mu\, m^\nu\, l^\kappa\, n^\lambda\,=\,0,\\
&& \Psi_2\,=\,-\,C_{\mu\nu\kappa\lambda}\,l^\mu\, m^\nu\, n^\kappa\, \overline{m}^\lambda\,=\,2C_2v^3,\\
&&\Psi_3\,=\,C_{\mu\nu\kappa\lambda}\,n^\mu\, \overline{m}^\nu\, n^\kappa\, l^\lambda\,=\,0,\\
&&\Psi_4\,=\,C_{\mu\nu\kappa\lambda}\,n^\mu\, \overline{m}^\nu\, n^\kappa\,\, \overline{m}^\lambda\,=\,0.
\end{eqnarray}
This result means that the solution has  Petrov classification $D$, the same classification as the empty space solutions of Schwarzschild and Kerr, and this may be understood as a confirmation that empty space solutions belong to the Petrov-type $D$. On the other hand, the axisymmetric solutions may belong to various Petrov classes \cite{treves:1980pet}, and what is most interesting, the solutions of Lewis and van Stockum, which are also non-asymptotically flat, belong to either  class $I$ or class $II$. Only Lewis and van Stockum solutions that are reducible to the Weyl class belong to type $D$. In spite of that, Petrov $D$ solutions have already been extensively studied in \cite{Kinnersley:1969zza}, although some shortcomings has been pointed up \cite{Edgar:2008hy,Ferrando:2013kwa}. In any case, the metric here presented in this form has never been studied with the proposal of this article.

At least we can calculate the Kretschmann scalar
\begin{equation}
K=R_{\mu\nu\kappa\lambda}R^{\mu\nu\kappa\lambda}=96C_2^2v^6,
\end{equation}
which is divergent in the limit $r\to \infty$, as $v$ is a quadratic function on $r$. This divergence has a profound meaning in the discussion that follows. We believe that it acts as a source of the gravitational field, and consequently mantains some resemblance to the Scharzschild solution.

\section{Light propagation\label{S3}}
We use the Lagrangian
\begin{equation}
\mathcal{L}=\frac{1}{2}\left[-\frac{u}{v^2}\,c^2\dot t^2+\frac{1}{v^4}\dot r^2+\frac{1}{uv^2}\dot z^2+r^2\dot\phi^2\right],
\end{equation}
where the dot represents the derivative respective to a proper time parameter $\tau$. The Lagrangian is independent of $t$ and $\phi$, and thence we obtain the conserved energy, $\mathcal{E}$, and the angular momentum, $\ell$, so that
\begin{equation}\label{charges}
\frac{u}{v^2}c^2\dot t =\mathcal{E}\qquad\mbox{and}\qquad r^2\dot\phi=\ell.
\end{equation}
From the line element, a  first integral of the equations of motion is obtained
\begin{equation}
-\frac{u}{v^2}\,c^2\dot t^2+\frac{1}{v^4}\dot r^2+\frac{1}{uv^2}\dot z^2+r^2\dot\phi^2=-c^2,
\end{equation}
which, with the conserved quantities, permits us to write
\begin{equation}\label{energy}
-\frac{v^2}{u}\frac{\mathcal{E}^2}{c^2}+\frac{\dot r^2}{v^4}+\frac{\dot z^2}{uv^2}+\frac{\ell^2}{r^2}=-c^2,
\end{equation}
which encodes the conservation of energy and momentum of a moving particle. This expression permits us to study the paths of particles and  light in the geometry. As a reference, we consider a light ray in the plane space.
\subsection{plane space}
For a light ray, $ds^2=0$, and the right hand side of (\ref{energy}) is zero. On the other hand, the proper time $\tau$ is understood as an affine parameter only. If $u=v=1$, the metric describes a plane space, and from (\ref{energy}) we obtain 
\begin{equation}\label{plane}
-\frac{\mathcal{E}^2}{c^2}+\dot r^2+\dot z^2+\frac{\ell^2}{r^2}=0.
\end{equation}
Imposing the constraint $z=\alpha r+\beta$, where $\alpha$ and $\beta$ are constants, we integrate (\ref{plane}) to obtain
\begin{equation}\label{int_plane}
\frac{\mathcal{E}^2}{\ell^2c^2}r^2-\frac{\mathcal{E}^4}{\ell^2c^4\big(1+\alpha^2\big)}\tau^2=1.
\end{equation}
Although $\tau$ is only a parameter and not the physical time, we can interpret the above ratio such that $\tau$ and $r$ are in a relativistic ``light cone'', in accordance with special relativity. On the other hand, looking at light from a direction parallel to the $z$ axis, so that $r$ and $\phi$ are constants, we obtain, as expected,  $z=\frac{\mathcal{E}}{c}\tau$. The same occurs fixing by $z$ and $\phi$ and varying $r$, and the relationship between the affine parameter and the coordinate is of course linear. Another aspect that must be considered is the pathway of light rays in the geometry. In the plane space-time, light rays are expected to travel in straight lines. In order to confirm this, we make $z=\alpha r+\beta$ and $r=r(\phi)$, and use the conservation laws (\ref{charges}) in (\ref{plane}) to obtain
\begin{equation}\label{plane_line}
\big(1+\alpha^2\big)\ell^2\frac{r^{\prime\,2}}{r^4}=\frac{\mathcal{E}^2}{c^2}-\frac{\ell^2}{r^2},
\end{equation}
where the prime denotes a $\phi$ derivative, whose integration yields
\begin{equation}\label{straight_line}
\frac{1}{r}=\frac{\mathcal{E}}{c\,\ell}\sin\frac{\phi}{\sqrt{1+\alpha^2}}.
\end{equation}
This equation represents a straight line distant $c\,\ell/\mathcal{E}$ from the origin of the coordinate system. The parameters $\alpha$ and $\beta$ are irrelevant for  geometrical interpretation, as the plane space is isotropic. Now we will consider the deviation from the linearity of light rays in curved space.
\subsection{curved space}
The choice  $z=\alpha r+\beta$ in equation (\ref{energy}) with the right hand side set to zero enables us to obtain
\begin{equation}\label{curved_light}
\dot r^2=\frac{v^4u}{u+\alpha^2v^2}\left(\frac{v^2}{u}\frac{\mathcal{E}^2}{c^2}-\frac{\ell^2}{r^2}\right).
\end{equation}
As the equation is too complex, we look for an approximate solution in order to obtain a comprehension of the effect of the geometry in the deviation from linearity in light rays.  The plane geometry is obtained when  $C_1=C_4=1$ and $r=z=0$, then we also set  $\beta=0$ and expand the geometrical elements of (\ref{curved_light}) in a McLaurin series, obtaining 
\begin{eqnarray}\label{curve_appr}
&&\dot r=\pm\sqrt{\frac{v^4u}{u+\alpha^2v^2}\frac{1-\frac{\mathcal{E}^2}{\ell^2c^2}\frac{v^2r^2}{u}}{1-\frac{\mathcal{E}^2}{\ell^2c^2}r^2}}\sqrt{\frac{\mathcal{E}^2}{c^2}-\frac{\ell^2}{r^2}}\nonumber\\
&&\dot r \left[1-\frac{C_3}{2}\frac{\alpha^3}{1+\alpha^2}r-\left(\frac{2+\alpha^2+2\alpha^4}{1+\alpha^2}C_2-\frac{\big(4+\alpha^2\big)\alpha^6}{8\big(1+\alpha^2\big)^2}C_3^2\right)r^2\right]\approx\frac{1}{\sqrt{1+\alpha^2}}\sqrt{\frac{\mathcal{E}^2}{c^2}-\frac{\ell^2}{r^2}}.
\end{eqnarray}
It is important to discuss the meaning of the above expansion, in order to mantain confidence in the result. The expansion has been carried out for the geometrical terms only, namely the functions $u$ and $v$ that come from the metric tensor. In fact, an expansion around $r=0$ of (\ref{curved_light}) would have no meaning, because of the divergence at this very point. Thus the expansion has been carried out in order not to affect the singularity and to obtain an equation that represents the propagation of  light  in a space-time which is approximately identical to the flat space in the vicinity of the point $r=z=0$ of the original metric. Of course, as the singularity has not been affected by the expansion, it remains in the final equation and the more terms we add to the expansion, the greater the effects of the singularity in the almost-flat metric. As we have just replaced one space-time with another, we are not restricted to the region around $r=0$ when we integrate (\ref{curve_appr}). This can be seen from the flat space solution (\ref{int_plane}), where the $r$ coordinate has a minimum value of $\frac{\ell c}{\mathcal{E}}$, and this condition remains valid for (\ref{curve_appr}).

Now, we integrate equation (\ref{curve_appr}), which represents the motion of a light ray in a approximately flat space that  recovers the flat space equation (\ref{plane}) if $C_2=C_3=0$. Using adimensional variable $x=\frac{\mathcal{E}}{c\ell}r$, we obtain from  (\ref{curve_appr})
\begin{equation}
\frac{\dot x\,x}{\sqrt{x^2-1}}\left[1-\frac{C_3}{2}\frac{\alpha^3}{1+\alpha^2}\frac{\ell c\,x}{\mathcal{E}}-\left(\frac{2+\alpha^2+2\alpha^4}{1+\alpha^2}C_2-\frac{\big(4+\alpha^2\big)\alpha^6}{8\big(1+\alpha^2\big)^2}C_3^2\right)\left(\frac{\ell c\,x}{\mathcal{E}}\right)^2\right]=\frac{1}{\sqrt{1+\alpha^2}}\frac{\mathcal{E}^2}{c^2\ell}
\end{equation}
and consequently
\begin{eqnarray}\label{light_int}
1&-&\frac{\mathcal{E}^2}{c^2\ell \sqrt{1+\alpha^2}}\frac{\tau}{\sqrt{x^2-1}}=\nonumber\\
&=&\frac{C_3}{4}\frac{\alpha^3}{1+\alpha^2}\frac{\ell c}{\mathcal{E}}\left(x+\frac{\ln\big(x+\sqrt{x^2-1}\big)}{\sqrt{x^2-1}}\right)+\left(\frac{2+\alpha^2+2\alpha^4}{1+\alpha^2}C_2-\frac{\big(4+\alpha^2\big)\alpha^6}{8\big(1+\alpha^2\big)^2}C_3^2\right)\left(\frac{\ell c}{\mathcal{E}}\right)^2\frac{x^2+2}{3}
\end{eqnarray}
As the flat space-time is recovered at  $C_2=C_3=0$, and as the right hand side of (\ref{light_int}) is zero in this limit, the right hand side of equation  (\ref{light_int}) measures the difference between the points that could be reached in the flat space and in the curved space using the same parametrization. This means that a light ray reaches a point at an equal distance $r$ at different values of the affine parameter $\tau$ depending on the curvature of the space. The sign of the constants $C_2$ and $C_3$ defines whether the difference of the point reached in the curved space can be reached either at a greater or at a smaller value of $\tau$. In the simplest situation where the light ray moves in the pure radial direction, such that $\alpha=0$, the difference $\Delta\tau$ is just 
\[ \Delta \tau=\frac{2\,C_2}{3} \frac{\mathcal{E}}{\ell^2 c}\big(x^2+2\big)\sqrt{x^2-1}, \]
and we see that what defines whether the difference is either positive or negative is the sign of $C_2$. In other words, the sign of this constant decides whether the distance in the curved space needs either more or less time to be crossed.

We now study the deviation of the light ray in curved space from the straight line observed in plane space. Using that $r=r(\phi)$ and the change of variable $x=\frac{\mathcal{E}}{c\ell}r$ , it is obtained around $r=0$ that 
\begin{equation}
\frac{ x'}{x\sqrt{x^2-1}}\left[1-\frac{C_3}{2}\frac{\alpha^3}{1+\alpha^2}\frac{\ell c\,x}{\mathcal{E}}-\left(\frac{2+\alpha^2+2\alpha^4}{1+\alpha^2}C_2-\frac{\big(4+\alpha^2\big)\alpha^6}{8\big(1+\alpha^2\big)^2}C_3^2\right)\left(\frac{\ell c\,x}{\mathcal{E}}\right)^2\right]=-\frac{1}{\sqrt{1+\alpha^2}},
\end{equation}
where the prime denotes the $\phi$ coordinate derivative and the minus sign on the right hand side gives the straight line in the plane limit. The integration gives
\begin{equation}
\arctan\frac{1}{\sqrt{x^2-1}}-\frac{\phi}{\sqrt{1+\alpha^2}}=\frac{C_3}{2}\frac{\alpha^3}{1+\alpha^2}\frac{\ell c}{\mathcal{E}}\ln\big(x+\sqrt{x^2-1}\big)+\left(\frac{2+\alpha^2+2\alpha^4}{1+\alpha^2}C_2-\frac{\big(4+\alpha^2\big)\alpha^6}{8\big(1+\alpha^2\big)^2}C_3^2\right)\left(\frac{\ell c}{\mathcal{E}}\right)^2\sqrt{x^2-1}.
\end{equation}
In order to understand the effect of the curvature, we expand the above series around $x=1$, and obtain
\begin{equation}
\frac{\phi}{\sqrt{1+\alpha^2}}=\frac{\pi}{2}-\sqrt{2(x-1)}\left[1-\frac{C_3}{2}\frac{\alpha^3}{1+\alpha^2}\frac{\ell c}{\mathcal{E}}-\left(\frac{2+\alpha^2+2\alpha^4}{1+\alpha^2}C_2-\frac{\big(4+\alpha^2\big)\alpha^6}{8\big(1+\alpha^2\big)^2}C_3^2\right)\left(\frac{\ell c}{\mathcal{E}}\right)^2\right].
\end{equation}
At $\phi=\frac{\pi}{2}$, we have the minimum distance between the light ray and the origin of the system or coordinates. We see the effect of the curvature of the space in the terms that depend on $C_2$ and $C_3$. Similar to what occurs with the affine parameter, the sign of  $C_2$ and $C_3$ determines whether the light ray on the curved space will deviate in one direction or another. In order to understand this behavior, let us consider the flat space solution (\ref{straight_line}) with a small variation $\delta$ 
\begin{equation}
\delta r=r\left(\frac{\pi}{2}\right)-r\left(\frac{\pi}{2}+\delta\right)=|\delta|.
\end{equation}
Of course, as $\phi=\frac{\pi}{2}$ is a minimum, every deviation, regardless of its sign, increases the $r$ coordinate. If the curvature makes the deviation larger than the plane space, the light ray will bend to become farther from the base line, where $\phi=0$. On the other hand, if the change $\delta r$ caused in the curved space is smaller than the deviation expected in the flat space, then the light ray will bend in the opposite direction and become closer to the base line $\phi=0$. As an example, if $\alpha=0$, and the light ray is parallel to the $z=0$ plane, we obtain

\begin{equation}
\phi=\frac{\pi}{2}-\sqrt{2(x-1)}\left[1-2\,C_2\left(\frac{\ell c}{\mathcal{E}}\right)^2\right].
\end{equation}
When $C_2<0$, the curved space makes $|\delta| $ greater than in the curved space, and in this situation the light ray will be farther than the base line $\phi=0$ line contained in the $z=0$ plane. This effect may be understood as if an anti-gravitational mass were contained in the origin of the coordinate system. On the other hand, when $C_2>0$ the light ray draws nearer the $z=0$ plane after passing near the origin, so that the distance between them diminishes after this point.

\section{orbital motion\label{S4}}
We proceed as in the light ray case, and consider a Newtonian planetary motion before studying the curved case.
\subsection{Newtonian theory}

In order to have a standpoint to study the curved case, we now tackle the well-known equations of the Newtonian planetary motion. The orbital motion of massive objects around a massive source of gravity is described by \cite{Stephani:1990gre}
\begin{equation}\label{rho_newton}
s''+s=\frac{MC^2}{\ell^2},
\end{equation}
where $s=1/r$, the prime denotes an angular derivative and $M$ is the mass of the source. Using the conserved angular momentum, we integrate (\ref{rho_newton}) and then change the variable back to $r$, obtaining the energy relationship
\begin{equation}\label{energy_newton}
\dot r^2=\frac{2Mc^2}{r}-\frac{\ell^2}{r^2},
\end{equation}
so that the right hand side of (\ref{energy_newton}) is interpreted as minus the potential. We can also study the force that acts on the particle using the time derivative of (\ref{energy_newton})
\begin{equation}
\ddot r=\frac{1}{r^3}\left(\ell^2-Mc^2r\right).
\end{equation}
The stable point of the dynamic system is obtained at $\dot r=0$, and for this case it is
\begin{equation}
r_0=\frac{\ell^2}{2Mc^2}.
\end{equation}
 At this very point, the potential is either a maximum or a minimum, so that the non-zero force that acts on the particle is given by
\begin{equation}
\ddot r=\frac{\ell^2}{2\,r_0^3}.
\end{equation}
This result is important as a guarantee that the particle will not escape from orbit. In order to determine whether $r_0$ is a maximum or a minimum of the potential, we calculate
\begin{equation}
\frac{d\ddot r}{dr}=\frac{2Mc^2r-3\ell^2}{r^4},\qquad\mbox{so that}\qquad \frac{d\ddot r}{dr}(r_0)=-\frac{2\ell^2}{r_0^4}.
\end{equation}
The second derivative at the point is negative, consequently the second derivative of the potential is positive, and $r_0$ is a minimum of the potential. Thence the particle oscillates around $r_0$ for slightly higher energy than the minimum  potential, and its movement is elliptical, as in the solutions of (\ref{rho_newton}).

\subsection{curved space}

In this section, we discuss  whether closed orbits are possible in the proposed space-time (\ref{metric}), but do not calculate them explicitly. We set $z=0$ in (\ref{energy}), and obtain the equations that govern the radial dynamics of the motion
\begin{eqnarray}\label{curved_energy}
\dot r^2&=&v^4\left[\,\frac{\mathcal{E}^2}{c^2}v^2-c^2-\frac{\ell^2}{r^2}\,\right]\\
\ddot r&=&v^3\,r\left[6\,C_2\frac{\mathcal{E}^2}{c^2}v^2+\frac{\ell^2v}{r^4}-4C_2\left(\frac{\ell^2}{r^2}+c^2\right)\right]
\end{eqnarray}
The equation $\dot r^2=0$ can be satisfied for $v=0$ if $C_2<0$. In the case of positive $C_2$, we isolate $v^2$ from $\dot r=0$ and substitute its value in $\ddot r$ obtaining
\begin{equation}
\ddot r\left(\dot r^2=0\right)=\frac{v^4\ell^2}{r^3}\left[1+2C_2\frac{\mathcal{E}^2}{\ell^2c^2}v\,r^4\right].
\end{equation}
The polynomial of sixth order inside the brackets can be solved in terms of a third order polynomial and it has at least one real root. Then the sixth order polynomial may have real solutions, depending on the values of the parameters. This proves that there is at least one equilibrium point in the system for each sign of $C_2$ without any approximation on $r$. In order to decide if $r_0$ is a maximum or a minimum, we write $\ddot r=v^3rf$, so that $f$ is the expression inside brackets in the definition of $\ddot r$ in (\ref{curved_energy}). Using this notation, we obtain
\begin{equation}\label{sec_curved}
\frac{d\ddot r}{dr}=v^3r\,f\left(\frac{3}{v}\frac{dv}{dr}+\frac{1}{r}+\frac{1}{f}\frac{df}{dr}\right).
\end{equation}
In order to have a minimum of the potential, we know from the Newtonian case that the sign of (\ref{sec_curved}) must be negative. From the positivity of $\ddot r$, it follows that $f(r_0)>0$, and the only way to have a negative sign in (\ref{sec_curved}) comes from the derivative or $f$. As (\ref{curved_energy}) shows, the potential has a singular value at $r=0$, and then it is highly positive at this point. For higher values of $r$, the polynomial dominates and changes the sign of the potential to a negative value. If the derivative of $\ddot r$ is positive and denotes a maximum of the potential, its value  is finite and greater than the value of the potential at a point close enough to $r=0$. Then, there is necessarily a minimum between this maximum and $r=0$. As the derivative has a term which depends on $\frac{1}{r^5}$ with  the negative sign, we conclude that there is a minimum there and consequently there are closed orbits in the metric. The complete characterization of these orbits and the values in the parameters that generate them is not of our interest here, the proof of its existence is enough to qualitatively characterize the metric. If there were no closed orbits, this model would be of almost no use in gravitation.

\section{the newtonian limit\label{S5}}
 The metric (\ref{metric}) is flat if $z=r=0$ and $C_1=C_4=1$. The other integration constants can be determined by using a weak field approach, so that the metric must give the approximate Newtonian gravity  when the gravitational field is weak. Considering that the geodesic equation is given in terms of the proper time derivatives $\dot x^\mu$ by
\begin{equation}
\ddot{x}^\mu=-\Gamma^\mu_{\nu\lambda}\dot{x}^\nu \dot{x}^\lambda,
\end{equation}
and that the gravitational field is generated by static particles, so that $x^\mu=(c,\,0,\,0,\,0)$, we get
\begin{equation}
\ddot{x}^\mu=-\Gamma^\mu_{00}c^2.
\end{equation}
In the weak field approach, the metric tensor $g_{\mu\nu}$ is a correction of the the Minkowski metric tensor $\eta_{\mu\nu}$ so that
\begin{equation}
g_{\mu\nu}=\eta_{\mu\nu}+h_{\mu\nu}.
\end{equation}
Using  $\eta_{\mu\nu}$ to move the indices and considering $h_{\mu\nu}$  time independent, we obtain
\begin{equation}
\ddot{x}^\mu=-\frac{c^2}{2}\eta^{\mu\nu}\partial_\nu h_{00}.
\end{equation}
On the other hand, the movement of a particle due to a gravitational potential $\Phi$ is given by
\begin{equation}
\bm\ddot{r}=-\nabla\Phi.
\end{equation}
In the specific case of a gravitational potential 
\begin{equation}
\Phi=\Phi_0+\phi,
\end{equation}
where $\Phi_0$ is a constant, we obtain
\begin{equation}\label{pfraco}
\phi=-\frac{c^2}{2}h_{00}.
\end{equation} 
Until now, we have followed the  usal procedure. Now, we can proceed to the specific case of the weak field generated by (\ref{metric}). Considering the axial symmetry of the model, we suppose that in the Newtonian limit the gravitational field is generated by a massive ring. Analogously as the electric potential generated by a charged ring \cite{Ciftja:2009ecr}, the gravitational potential generated by a ring of radius $R$ and mass $m$ is
\begin{equation}
\Phi(r,\,z)=-\frac{2Gm}{\pi}\frac{1}{\sqrt{(r+R)^2+z^2}}\;\textrm{K}\!\left(\frac{4Rr}{(r+R)^2+z^2}\right),
\end{equation}
where K denotes an elliptic integral and $G$ is the Newton constant. Setting $r=0$, we have around $z=0$ that
\begin{equation}
\Phi(r,\,z)=\frac{mG}{R}\Big(-1+\frac{z^2}{2R^2}\Big),
\end{equation}
and then (\ref{pfraco}) implies
\begin{equation}
C_2=\frac{mG}{4c^2R^3}\qquad\textrm{and}\qquad C_3=0.
\end{equation}
It is important to note that the expansion of the potential (\ref{pfraco}) around  $r=0$ generates a series whose first term depends on the first order in $r$, and that the expansion of the metric element around $r=0$ produces a first term of second order in $r$. Then, the metric has a weaker dependence on the radial coordinate than the massive ring. We can understand that the calculated metric may  be generated by a axially symmetric structure, but not necessarily a ring.  On the other hand,  the massive ring is located at infinity, and this can explain why the field generated by such an object is weaker than the gravitational field generated by a finite one. The exact form of this ideal object is  a subject for future research.

On the other hand, this analysis shows that, if the $r\to\infty$ is a singular point, it is indeed naked, because $C_2>0$ and the metric has no singular point at a finite value of $r$. As discussed in the introduction, this space has the interesting effect of producing a gravitational field that increases with the increasing  the value of the coordinate. This means that a massive point in a freely fall towards the $r\to\infty$ describes an increase in the distances between the massive points. This seems an interesting topic for future application on cosmological models, where inflation and dark matter are the most studied models which describe such an effect.

\section{conclusion\label{S6}}
In this article we have described a space-time which is empty and axisymmetric.  Its Petrov classification is $D$, the same of the spherical empty space solutions of Kerr and Schwarzchild, something that establishes a connection among these solutions. This geometry cannot be transformed into the well-known Weyl type of cylindrical symmetry, and cannot also be put into an isotropic coordinate system like Schwarzschild geometry. The moral of this case is that this curious object has potential importance as a model for possible applications in inflationary cosmology and also because it is a simple solution, and physics needs simple and well-known objects in order to model more complex systems. Future directions of research are many and varied. From the mathematical standpoint, it is interesting to determine whether there are space-time singularities in the metric. There are the obvious applications in cosmology and there are also other possibilities for discovering  new exact solutions. Solutions involving electric charge or angular momentum are the most obvious examples in the latter direction.

\section*{acknowledgements}
Sergio Giardino receives a financial grant from Capes for his research.
%
%
%
%


\end{document}